\documentclass[12pt]{article}
\usepackage{epsfig}
\begin{document}
\title{Dimers and the Ising model}
\author{ Jerzy Cis{\l}o \\
Institute of Theoretical Physics  \\
University of Wroc{\l}aw  \\
pl. M.Borna 9, 50-205 Wroc{\l}aw, Poland  \\
cislo@ift.uni.wroc.pl
}
\maketitle
\date{ }
\begin{abstract}
We present a innovative relationship between ground states of the Ising model
and dimer coverings which sheds new light on the Ising Models with highly degenerated ground states and enables one to construct such models. Thanks to this relationship we also
find the generating function for dimers
as the appropriate limit of the free energy per spin for the Ising model.
\end{abstract}
\noindent {\bf 1. Dimers.}
Dimers are objects connecting two neighboring
sites of the lattice. We consider a lattice which can be covered
with dimers in such a way that each site of the lattice belongs to exactly
one dimer. How to calculate the number of such coverings for a given lattice was shown in 1961 by Kasteleyn, Fisher and Temperley in 1961 [1-3]. The connection between the Ising
model on the planar lattice and the dimer problem is well known.
In 1961, Fisher [4] gave a general rule connecting the partition function
for the Ising model with a generating function for configurations of dimers
on an appropriately constructed lattice. For the Ising model on the honeycomb lattice, for instance, the dimers on 3-12 lattice are considered. Moreover, the Ising model---as well as the model of free fermions---are
connected with the dimer problem on a nonplanar lattice [5]. One more technique for the dimer problem
was proposed by Percus [6] in 1969. In 2006 Cimasoni and Reshetikhin [7] also considered a relation between the dimer problem and spin structures. Besides, the notion of dimer occurs in pure mathematics [8].

\noindent {\bf 2. From the Ising model to the dimer problem.}
In this section we shall present a new, different from the ones described by the above-mentioned authors, relationship between the Ising model and the problem of dimers. We shall show that the doubled number of dimer
coverings for a given lattice is equal to the degeneration of the ground state of
the Ising model on a lattice dual to the one covered with dimers. 

Let us begin with the description of our construction. Let $L$ be any
planar lattice for which the dual lattice $L_{d}$ can be covered
with dimers. Let us cover the lattice $L_{d}$ with dimers in one of the possible ways
(standard configuration).

To every edge crossed by a dimer we assign the positive constant of interaction $E$,
and to the remaining edges, the negative constant of interaction $-E$.

The energy of the interaction of two neighboring spins is defined by the product
of these spins and the constant of interaction. The degeneracy of the ground state
equals the number of spin configurations with the lowest energy.

Our main observation is that the degeneracy of the ground state of our Ising model is
equal to half of the number of dimer coverings of the lattice $L_{d}$.

To prove the above statement, let us note that we get the smallest energy when
exactly one edge of each elementary polygon of the lattice $L$ has the positive energy $E$.

Each configuration of this type corresponds to one dimer covering of the lattice $L_{d}$.
Conversely, each dimer covering corresponds to two spin
configurations with the lowest energy.
To see that, we fix one spin and systematically define the neighboring ones:
if the edge is positive--- or negative but crossed--- the next spin has the same sign;
otherwise the next spin has the opposite sign.

Obviously, we
can fix the first spin in two ways. Hence, the number of dimer coverings reads

\begin{equation}
Z_d=\frac{1}{2}
 \lim_{T\rightarrow 0^+} Z(T) e^{M E / k T},
\end{equation}
where $Z$ is the statistical sum of the Ising model on the lattice $L$,
$ME$ is a minimal energy of the system, and $M$ is the number of edges of the lattice
$L$ minus number of dimers.

\noindent {\bf 3. Example.}
Here, we present an application of the procedure described above.
We shall find the generating function for the dimers on the chessboard lattice.

Let us consider the chessboard Ising model with constants of interaction
$E_{1}, E_{2}, E_{3} <0, E_{4}>0$ (see Fig.1).

We change the signs of $E_{2}$ and $E_{4}$ in every other column so that the edges crossed
by dimers of the standard configuration have positive interaction constant (Fig.1).
This transformation does not change the statistical sum.

Let $E_1=-E-L_1kT,\; E_2=-E-L_2kT, \; E_3=-E-L_3kT,\; E_4=E+L_4kT$.

Then
\begin{equation}
\frac{1}{2}\lim_{T\rightarrow 0^+} Z(E_1,E_2,E_3,E_4)
e^{ME/kT}=Z_d(x_1,x_2,x_3,x_4),
\end{equation}
where $Z(x_1,x_2,x_3,x_4)$
is the generating function for the dimer problem on the chessboard lattice
in the variables $x_i=\exp(2L_i-L_1-L_2-L_3-L_4)$, $M=3N/4$, and
$N$ is the number of spins.

In 1951 Utiyama obtained the formula for the free energy per spin for
the Ising model on the chessboard lattice [9]:
$$
  f(T)=-kT \lim_{N \rightarrow \infty } \frac{\ln Z(T)}{N} =
$$
\begin{equation}
= -kT \ln 2 - \frac{kT}{16 \pi^{2} } \int_{0}^{2\pi} \int_{0}^{2\pi}
 \ln \frac{1}{2} ( 1+C_{1}C_{2}C_{3}C_{4}+S_{1}S_{2}S_{3}S_{4}+
\end{equation}
$$
+(S_{1}S_{2}+S_{3}S_{4}) \cos \phi -(S_{1}S_{4}+S_{2}S_{3}) \cos \theta +
$$
$$
 +S_{2}S_{4} \cos(\phi+\theta) +S_{1}S_{2}
\cos(\phi-\theta) ) d\theta d\phi ,
$$
where $C_{i}=\cosh (-2E_{i}/kT)$, and $S_{i}=\sinh (-2E_{i}/kT)$.

While calculating the thermodynamic limit, we suppose that the lattice grows in the same way in two directions.

Using this result we find
\begin{equation}
\Psi(x_{1},x_{2},x_{3},x_{4}) =
\lim_{N \rightarrow \infty } { \ln Z_d(x_{1},x_{2},x_{3},x_{4}) \over N}=
\frac{3}{4} - \lim_{T \rightarrow 0^+} \frac{ f(T)}{kT} =
\end{equation}
$$
={1 \over 16 \pi^{2} } \int_{0}^{2\pi} \int_{0}^{2 \pi} \ln (
x_{1}^{2}+x_{2}^{2}+x_{3}^{2}+x_{4}^{2} +
$$
$$
+2(x_{3}x_{4}-x_{1}x_{2}) \cos \phi +2(x_{2}x_{3}-x_{1}x_{4}) \cos \theta -
$$
$$
 -2x_{1}x_{3} \cos (\phi+\theta) +2x_{2}x_{4} \cos
(\phi-\theta) ) d\theta d\phi =
$$
$$
= \frac{1}{8 \pi} \int_{0}^{2\pi}
\log \max ( x_1^2+x_2^2-2x_1x_2 \cos \phi, ~ x_3^2+x_4^2 + 2 x_3 x_4 \cos \phi ) \; d\phi.
$$
If $\max(x_1,x_2,x_3,x_4) \geq (x_1+x_2+x_3+x_4)/2$,
then the result is especially simple:
\begin{equation}
\Psi(x_{1},x_{2},x_{3},x_{4})= \frac{1}{2} \log \max(x_1,x_2,x_3,x_4).
\end{equation}

In this way it was possible to find the generating function for the dimer problem
without any reference to the general theory. Similar considerations may be repeated
for the triangular and honeycomb lattices.

Finally, it should be stressed that, for the purpose of simplicity and clarity,
we decided to neglect the boundary conditions. For large systems they do not seem to matter.
One might however easily take them into account in our reasoning:
it is enough to assume that all the boundary spins in the Ising model are of the same sign.
The influence of boundary conditions on the number of dimer coverings was described by Cohn,
Kenyon and Propp [10].
 
\begin{figure}
 \centering
 \includegraphics[width=7cm,height=7cm]{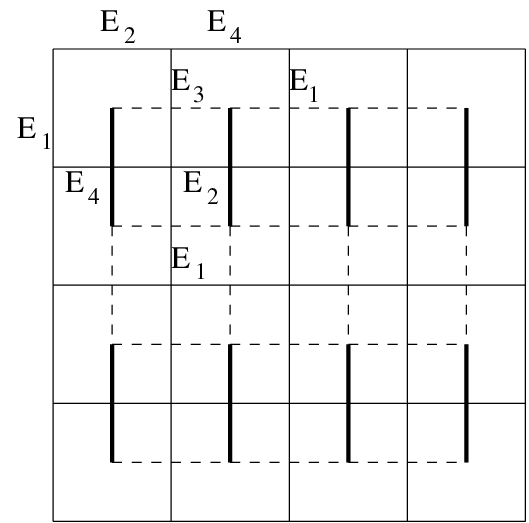}

 \caption{Fragments of lattices: $L$ (continues lines), $L_{d}$ (dotted lines),
  and standard configurations of dimers (bold lines).}
\end{figure}


\begin{thebibliography}{99}
\bibitem{1} P.W. Kasteleyn, Physica {\bf 27} (1961) 1209.
\bibitem{2} M.E. Fisher, Phys.Rev. {\bf 124} (1961) 1664.
\bibitem{3} H.N.V. Temperley, M.E. Fisher, Phil.Mag. {\bf 6} (1961) 1061.
\bibitem{4} M.E. Fisher, J.Math.Phys. {\bf 7} (1961) 1776.
\bibitem{5} P.W. Kasteleyn, J.Math.Phys. {\bf 4} (1963) 287.
\bibitem{6} J.K. Percus, J.Math.Phys. {\bf 10} (1969) 1881.
\bibitem{7} D. Cimasoni, N. Reshetikhin, (2006) arXiv:math-ph/0608070v2.
\bibitem{8} W. McCuaig, Electron.J. of Combin. 11 (2004) R79.
\bibitem{9} T. Utiyama, Prog.Theor.Phys. {\bf 6} (1951) 907.
\bibitem{10} H. Cohn, R. Kenyon, J. Propp, Journal of AMS {\bf 14} (2000) 297.
\end{thebibliography}
\end{document}